\documentclass[aps,prd,twocolumn,groupedaddress]{revtex4}
\usepackage{graphicx}
\usepackage{dcolumn}
\usepackage{bm}
\usepackage{amssymb,amsmath,latexsym,amsfonts}
\usepackage{amssymb}
\usepackage{graphicx}

\begin{document}

\title{Weakly interacting Bose-Einstein condensates in temperature-dependent
generic traps}
\author{E. Castellanos}
\email{ecastellanos@fis.cinvestav.mx}
\affiliation{Departamento de F\'{\i}sica, Centro de Investigaci\'on y Estudios Avanzados
del IPN\\
A. P. 14--740, 07000, M\'exico, D.F., M\'exico.}
\author{F. Briscese}
\email{fabio.briscese@sbai.uniroma1.it}
\affiliation{Istituto Nazionale di Alta Matematica Francesco Severi, Gruppo Nazionale di
Fisica Matematica, Citt\`a Universitaria, P.le A. Moro 5, 00185 Rome, EU.
and Dipartimento SBAI, Sezione di Matematica, Sapienza Universit\`a di Roma,
Via Antonio Scarpa 16, 00161 Rome, EU. }
\author{M. Grether}
\email{mdgg@hp.fciencias.unam.mx}
\affiliation{Facultad de Ciencias, Universidad Nacional Aut\'onoma de M\'exico \\
- 04510 M\'exico, DF, M\'exico.}
\author{M. de Llano}
\email{dellano@unam.mx}
\affiliation{Instituto de Investigaciones en Materiales, Universidad Nacional Aut\'onoma
de M\'exico, \\
A. P. 70-360, 04510 M\'exico, DF, M\'exico.}

\begin{abstract}
The shift in condensation temperature caused by interactions is studied up
to second order in the s-wave scattering length in a Bose-Einstein
condensate trapped in a temperature-dependent three-dimensional generic
potential. With no assumptions other than the mean-field approach and
semiclassical approximations it is shown that the inclusion of a
temperature-dependent trap improves the empirical values of the numerical
parameters compared to those obtained in previous reports on the temperature
shift.
\end{abstract}

\maketitle

\section{Introduction}

Since its theoretical prediction by Bose and Einstein \cite{Bose,Einstein} in
the 1920s until its laboratory observation from 1995 onwards with
magneto-optical traps\cite{Anderson,Davis,CC,CC1} Bose-Einstein condensation
(BEC) of dilute atomic gases has stimulated an enormous amount of related
work. Among the issues addressed one finds, e.g., mathematical questions
related to BEC\cite{Lieb}, diverse theoretical and heuristic aspects\cite%
{Dalfovo,S}, and as a viable tool for precision tests in gravitational
physics\cite%
{CastellanosCamacho,CastellanosCamacho1,Castellanos,r1,r2,elca,eli,eli2,CastellanosClaus}%
. 

The study of its associated thermodynamic properties is naturally also a
very pertinent aspect of BECs\cite{klu,Pethick,ueda,EM,ME}. Indeed, the
condensation temperature $T_{c}$, \emph{i.e.,} the critical temperature
below which a macroscopic quantum state of matter appears, has been the
subject of considerable discussion, see Ref.\cite{Dalfovo,JOA} and refs.
therein. In particular, the influence of interparticle interactions on $T_{c}
$ turns out to be a deep nontrivial matter, see e.g. Refs.\cite%
{yuka0,smith,smith1}.

Interboson interactions produce a shift $\Delta T_{c}/T_{c}^{0}\equiv
(T_{c}-T_{c}^{0})/T_{c}^{0}$\ in the condensation temperature $T_{c}$ with
respect to that of the ideal noninteracting case $T_{c}^{0}$ in the
thermodynamic limit. For instance, the
contributions to $\Delta T_{c}/T_{c}^{0}$ due to interactions in a uniform
dilute gas originate in the fact that the associated many-body system is
affected by long-range critical fluctuations rather than by purely
mean-field (MF) considerations\cite{JOA,Baym,Hol}. However, it is generally
accepted that $\Delta T_{c}/T_{c}^{0}$ for this system behaves like $%
c_{1}\delta +(c_{2}^{\prime }\ln \delta +c_{2}^{\prime \prime })\delta ^{2}$%
, with the dimensionless variable $\delta \equiv \rho ^{1/3}a$ where $\rho $
is the corresponding boson number density, $a$ the $S$-wave two-body
scattering length\cite{Baym} related to the pair interaction, and the $c_{i}$%
's are dimensionless constants. A good fit\cite{yuka0} gives $c_{1}\simeq
1.32$, $c_{2}^{\prime }\simeq 19.75$ and $c_{2}^{\prime \prime }\simeq 75.7$.

It is noteworthy that these ideas can be extended to more general traps\cite%
{zobay,zobay1,zobay2} in which the relative shift $\Delta T_{c}/T_{c}^{0}$\
on the condensation temperature explicitly exhibits a sensitive
trap-dependence. This extension to \textit{generic} traps allows summarizing
the corrections on $\Delta T_{c}/T_{c}^{0}$ as function of a simple index
parameter describing the trap \textit{shape}.

On the other hand, when interactions are considered for the more common 
\textit{harmonic} traps one finds a shift in $T_{c}$ up to second order in
the $S$-wave scattering length $a$ within the MF approach given by\cite%
{smith,smith1} 
\begin{equation}
\frac{\Delta T_{c}}{T_{c}^{0}}\simeq b_{1}(a/\lambda
_{T_{c}^{0}})+b_{2}(a/\lambda _{T_{c}^{0}})^{2}  \label{deltaTsuNONTUNIFORM1}
\end{equation}%
where 
\begin{equation}
k_{B}T_{c}^{0}=\hbar \omega \lbrack N/\zeta (3)]^{1/3}  \label{CTI}
\end{equation}%
(with $\zeta (3)\simeq 1.202$) is the condensation temperature associated
with the ideal system ($a=0$) in the thermodynamic limit \cite{Pethick}, and 
$b_{1}\simeq -3.426$\cite{fabio} while $b_{2}\simeq 11.7$\cite{smith1}.
Furthermore, these results seem to contrast with the results reported, e.g.,
in Refs.\cite{a7,AR} since, as was mentioned in Ref.\cite{smith}, the
well-known logarithmic corrections to (\ref{deltaTsuNONTUNIFORM1})\ are not
discernible within the error bars.

Note that from (\ref{deltaTsuNONTUNIFORM1}) $\Delta T_{c}$ is negative for
repulsive interactions \emph{i.e.}, $a>0$. The result (\ref%
{deltaTsuNONTUNIFORM1}) is in excellent agreement with laboratory
measurements of $\Delta T_{c}/T_{c}^{0}$\cite{b1,b2,b3,smith1} to first
order in $(a/\lambda _{T_{c}^{0}})$ but differs somewhat with data to second
order $(a/\lambda _{T_{c}^{0}})^{2}$. In Ref.\cite{smith}, high precision
measurements of the condensation temperature of the bosonic atom $^{39}K$
vapor in the range of parameters $N\simeq (2-8)\times 10^{5}$, $\omega
\simeq (75-85)Hz$, $10^{-3}<a/\lambda _{T_{c}^{0}}<6\times 10^{-2}$ and $%
T_{c}\simeq (180-330)nK$ have detected second-order effects in $\Delta
T_{c}/T_{c}^{0}$. The measured $\Delta T_{c}/T_{c}^{0}$ is well fitted by a
quadratic polynomial (\ref{deltaTsuNONTUNIFORM1}) with best-fit parameters $%
b_{1}^{exp}\simeq -3.5\pm 0.3$ and $b_{2}^{exp}\simeq 46\pm 5$ so that the
value $b_{2}\simeq 11.7$\cite{smith1} is strongly excluded by data. This
discrepancy between (\ref{deltaTsuNONTUNIFORM1}) and data may be due to
beyond-MF effects (see Ref.\cite{smith1}). Beyond-MF effects are expected to
be important near criticality, where the physics is often nonperturbative.
It is therefore sounds reasonable that a beyond-MF treatment might give a
correct estimation of $b_{2}$. However, this is not certain since beyond-MF
effects have been calculated in the case of \textit{uniform} condensates\cite%
{a7,a8} but are still poorly understood for \textit{trapped} BECs\cite%
{AR,c2,c3,c4,c5}. It thus seems that it is currently not possible to
ascertained whether the discrepancy between $b_{2}$ and $b_{2}^{exp}$ can be
explained in the MF context or arises from beyond-MF effects.

Nevertheless, the effect of interactions on the condensation temperature $%
T_{c}$ of a Bose-Einstein condensate trapped in a harmonic potential was
recently discussed\cite{fabio}. In the latter paper it was shown that,
within the MF Hartree-Fock (HF) and semiclassical approximations,
interactions among the particles produce a shift $\Delta
T_{c}/T_{c}^{0}\simeq b_{1}(a/\lambda _{T_{c}^{0}})+b_{2}(a/\lambda
_{T_{c}^{0}})^{2}+\psi \left[ a/\lambda _{T_{c}^{0}}\right] $ with $\lambda
_{T_{c}^{0}}\equiv (2\pi \hbar ^{2}/mkT_{c}^{0})^{1/2}$ the thermal
wavelength, and $\psi \left[ a/\lambda _{T_{c}^{0}}\right] $ a non-analytic
function such that $\psi \left[ 0\right] =\psi ^{\prime }\left[ 0\right]
=\psi ^{\prime \prime }\left[ 0\right] =0$ but $|\psi ^{\prime \prime \prime
}\left[ 0\right] |=\infty $. Therefore, with only the usual assumptions of
the HF and semiclassical approximations, interaction effects are
perturbative to second order in $a/\lambda _{T_{c}^{0}}$ and the expected
nonperturbativity of physical quantities at the critical temperature emerges
only at third order. Indeed, in Ref.\cite{fabio} an analytical estimation
for $b_{2}\simeq 18.8$ was obtained which improves the previous numerical
fit-parameter value of $b_{2}\simeq 11.7$ obtained in Ref.\cite{smith1}.
Even so, the value for $b_{2}$ obtained in Ref.\cite{fabio} still differs
substantially from the empirical value $b_{2}^{exp}\simeq 46\pm 5$\cite%
{smith}.

We mention that the temperature shift $\Delta
T_{c}/T_{c}^{0}$ induced by interparticle interactions obtained in Ref.\cite%
{fabio} seems to contradict, for instance, the result reported in Ref.\cite%
{AR} where the interaction induced temperature shift is estimated as 
\begin{equation}
\frac{\Delta T_{c}}{T_{c}^{0}}=b_{1}(a/\lambda _{T_{c}^{0}})+\Bigl(%
b_{2}^{\prime }+b_{2}^{\prime \prime }\ln (a/\lambda _{T_{c}^{0}})\Bigr)%
(a/\lambda _{T_{c}^{0}})^{2}
\end{equation}%
with $b_{1}\simeq -3.426$, $b_{2}^{\prime }\simeq -45.86$ and $b_{2}^{\prime
\prime }\simeq -155.0$ \cite{a7} (see also Ref.\cite{yuka0} for a
discussion). This result has been obtained using lattice simulations and a
technique based on a scalar field analogy, but is questionable (see
discussion in Ref. \cite{fabio}) besides being in striking contradiction to
the data.

It is thus clear that these results differ substantially from the
estimations obtained in Ref.\cite{fabio} and the results obtained here (see
below), but also conflict with the results obtained in Ref.\cite{smith1} as
well as experiment\cite{smith}. We therefore stress that before addressing
beyond-MF effects these facts suggest that MF effects might still be
well-understood and deserve further analysis.

In fact, in a recent paper\cite{malik} the use of an effective \textit{%
temperature-dependent trapping potential} was suggested in order to
calculate the condensation temperature of noninteracting systems; see also
Ref.\cite{llano} for a wide-ranging justification of $T$-dependent
Hamiltonians. Hence, it might be useful to explore this idea in the context
of the effects on the condensation temperature caused by interparticle
interactions.

These considerations pushed us into the novel terrain of $T$-dependent
Hamiltonians, and more specifically to $T$-dependent trapping potentials. We
note that this it is not the first time that such a terrain has been
reached, e.g., we find the employment of $T$-dependent dynamics in: a)
superconductivity in the work of Bogoliubov, Zubarev and Tserkovnikov, as
discussed by Blatt\cite{Blatt}; b) an explanation\cite{Malik} of the
empirical law in superconductors $H_{c}(T)=H_{c}(0)[1-(T/T_{c})^{2}]$ where $%
H_{c}(T)$ is the critical field at $T$; c) finite-$T$ behavior\cite%
{EM,ME,Linde,Linde1,Wei,Dolan} of a class of relativistic field theories
(RFTs) to address the question of restoration of a symmetry which at $T=0$
is broken either dynamically or spontaneously; d) the Wick-Cutkosky model%
\cite{Malik1} in an RFT; 5) legions of unidentified solar-emission lines\cite%
{Malik2}; e) QCD to explain \cite{Malik3,Malik4} the masses of different
quarkonium families and their deconfinement temperatures; and most recently,
as was mentioned above, f) in a comparative study\cite{malik} of the
experimental features of the Bose-Einstein condensates in several species of
bosonic atomic gases.

We thus examine the possibility of such $T$-dependent generic potentials in
order to analyze (or even improve upon) the value $b_{2}\simeq 18.8$
obtained in Ref.\cite{fabio} within the HF MF theory, and also to explore
its discrepancy with the empirical value $b_{2}^{exp}\simeq 46\pm 5$. For
all this we now entertain $T$-dependent generic traps $V(r,T)$.


\section{Mean field Hartree-Fock approximation}

Following Ref.\cite{fabio} we define the following semiclassical energy
spectrum in the MF HF approximation (see, e.g.,\cite{Dalfovo,Pethick}) 
\begin{equation}
E(p,r,g)=\epsilon (p,r)+2g\,n(r,g)  \label{energy interaction}
\end{equation}%
where $\epsilon (p,r)\equiv p^{2}/2m+V(r)$ with $V(r)$ the external
potential, $n(r,g)$ the spatial density of bosons, and $g\equiv 4\pi \hbar
^{2}a/m$, the parameter describing the interaction.

Moreover, the semiclassical condition allows approximating summations over
energy states by integrals, namely $\sum_{\mathbf{k,r}}\rightarrow \int
d^{3}rd^{3}p/(2\pi \hbar )^{3}$. Therefore, the number of particles $N$ in
three-dimensional space obeys the normalization condition\cite%
{Dalfovo,Pethick} 
\begin{equation}
N=N_{0}+\int \frac{d^{3}rd^{3}p}{(2\pi \hbar )^{3}}\left( \exp \left[ \frac{%
E(p,r,g)-\mu }{k_{B}T}\right] -1\right) ^{-1}  \label{semiclassicalNth1}
\end{equation}%
where $N_{0}$ is the number of particles in the ground state, $\mu $ the
corresponding chemical potential, and $k_{B}$ the Boltzmann constant.

At the condensation temperature $T_{c}$ we assume within MF theory that the
chemical potential $\mu $ is given by\cite{fabio} 
\begin{equation}
\mu _{c}(g)=2g\,n(r=0,g).  \label{numericalmuc}
\end{equation}%
Further assuming just above $T_{c}$ that in the ground state $N_{0}$\ is
negligible it follows that 
\begin{equation}
\begin{array}{ll}
N\pi \hbar ^{3}/2=\int drdpr^{2}p^{2}\left( \exp \left[ \frac{E(p,r,g)-\mu
_{c}(g)}{k_{B}T_{c}(g)}\right] -1\right) ^{-1} &  \\ 
\equiv \int d\Omega \Lambda \left[ \theta \right] &  \\ 
& 
\end{array}
\label{numericaltc1}
\end{equation}%
where 
\begin{equation}
\begin{array}{ll}
d\Omega \equiv drdpr^{2}p^{2}\qquad \Lambda \left[ \theta \right] \equiv %
\left[ \exp {\left[ \theta \right] }-1\right] ^{-1} &  \\ 
\theta \equiv \frac{\epsilon (p,r)+2\bar{n}(r,g)}{k_{B}T_{c}(g)}\qquad \bar{n%
}(r,g)\equiv n(r,g)-n(0,g). & 
\end{array}
\label{nbardefinition}
\end{equation}

From (\ref{numericaltc1}) we are able to extract, in principle, $T_{c}$ as a
function of the parameter $g$ describing interactions. Note that the
scattering length $a$ can be positive or negative, its sign and magnitude
depending crucially on the details of the atom-atom potential\cite{Dalfovo}.
However, a negative scattering length could lead to instabilities within the
system\cite{Pethick}, and finite-size effects could be important in this
situation due to the number of particles $N$ not being large enough\cite%
{Dalfovo}. Here, we restrict ourselves, as usual, to positive values of the
interaction parameter $g$ in order to compare our results with the reported%
\cite{smith} experimental data.

On the other hand, if $\Delta T_{c}$ is analytic in $g$ one can express the
relative shift in $T_{c}$ for small values of $g$ as follows 
\begin{equation}
\frac{\Delta T_{c}}{T_{c}^{0}}=\sum_{h=1}^{\infty }\frac{g^{h}}{h!}\frac{%
\partial _{g}^{h}T_{c}(g)}{T_{c}(g)}{\Huge |}{_{g=0}.}  \label{deltaTsuT0}
\end{equation}%
Note that $T_{c}(g=0)=T_{c}^{0}$ is by definition the $T_{c}$ for the
noninteracting system, given by (\ref{CTI}). Additionally, the expansion
coefficients can be expressed as%
\begin{equation}
\frac{\partial _{g}^{h}T_{c}(g)}{T_{c}(g)}{\Huge |}_{g=0}\equiv \frac{I_{h}}{%
\left( k_{B}T_{c}^{0}\lambda _{T_{c}^{0}}^{3}\right) ^{h}}  \label{Shift}
\end{equation}%
where the numerical factors $I_{h}$ depend on the external potential under
consideration and can be calculated explicitly.

This enables one to reexpress (\ref{deltaTsuT0}) as a power series in the
dimensionless interaction-dependent variable $a/\lambda _{T_{c}^{0}}$ 
\begin{equation}
\frac{\Delta T_{c}}{T_{c}^{0}}=\sum_{h=1}^{\infty }\frac{2^{h}I_{h}}{h!}%
\left( a/\lambda _{T_{c}^{0}}\right) ^{h}\equiv \sum_{h=1}^{\infty
}b_{h}\left( a/\lambda _{T_{c}^{0}}\right) ^{h}  \label{deltaTsuT4}
\end{equation}%
which defines the coefficients $b_{h}$. For an isotropic harmonic potential $%
V(r)\sim r^{2}$ the first two factors $I_{1}$ and $I_{2}$ are given
respectively by\cite{fabio} 
\begin{equation}
I_{1}=2\frac{\int d\Sigma \,\Lambda ^{\prime }\left[ u^{2}+v^{2}\right] Q%
\left[ v^{2}\right] }{\int d\Sigma \,\left( u^{2}+v^{2}\right) \Lambda
^{\prime }\left[ u^{2}+v^{2}\right] }  \label{I1}
\end{equation}%
\begin{equation}
\begin{array}{ll}
I_{2}=4\int \,d\Sigma \,\left[ \Lambda ^{\prime }\left[ u^{2}+v^{2}\right] S%
\left[ v^{2}\right] +\right. \Lambda ^{\prime \prime }\left[ u^{2}+v^{2}%
\right] \times &  \\ 
\left. \left[ Q\left[ v^{2}\right] -{\frac{1}{2}}[u^{2}+v^{2}]I_{1}\right]
^{2}\right] /\int d\Sigma \left( u^{2}+v^{2}\right) \Lambda ^{\prime }\left[
u^{2}+v^{2}\right] &  \\ 
& 
\end{array}
\label{I2}
\end{equation}%
where $d\Sigma \equiv dudvu^{2}v^{2}$, $Q[\alpha ]\equiv g_{3/2}\left[ \exp
(-\alpha )\right] -g_{3/2}\left[ 1\right] $, and $g_{\alpha
}[z]=\sum_{k=1}^{\infty }z^{k}/k^{\alpha }$ is the so-called Bose-Einstein
function\cite{Phatria}. Thus $S[\alpha ]\equiv \frac{3}{2}I_{1}Q[\alpha
]+\left( \alpha I_{1}-2Q[\alpha ]\right) g_{1/2}\left[ \exp (-\alpha )\right]
$ with $\alpha \equiv \left[ V(r)+2g\bar{n}(r,g)\right] /k_{B}T_{c}(g)$, see
Ref.\cite{fabio} for details.

Note that the assumptions used above lead to $b_{1}\simeq -3.426$ in
agreement with the experimental $b_{1}\simeq -3.5\pm 0.3$ obtained in Ref.%
\cite{smith}. Also, one gets $b_{2}\simeq 18.8$ which improves upon the
estimation of $b_{2}\simeq 11.7$ in Ref.\cite{smith1}. However, this value
still remains much smaller than the experimental estimation $%
b_{2}^{exp}\simeq 46\pm 5$ reported in Ref.\cite{smith}.

\section{$T$-dependent generic potentials and $T_{c}$}

Here we consider the following $T$-dependent generic potentials 
\begin{equation}
V(r,T)=\frac{m\omega ^{2}r^{2}}{2}\left[ 1+d\left( \frac{m\omega ^{2}r^{2}}{%
2k_{B}T}\right) ^{\beta /2}\right]  \label{pot1}
\end{equation}

\begin{equation}
V(r,T)=\frac{m\omega ^{2}r^{2}}{2}\Big(\frac{m\omega ^{2}r^{2}}{2k_{B}T}\Big)%
^{\delta /2}  \label{pot2}
\end{equation}%
for $T=T_{c}$ and with $d$, $\beta $, and $\delta $ dimensionless parameters.

\subsection{$T$-dependent generic potential with free parameters $d$ and $%
\protect\beta $}

Here we use the potential (\ref{pot1}) and find $b_{1}(d,\beta )$ from (\ref%
{Shift}) for $h=1$ as a function of $d$ and $\beta ,$which reads%
\begin{equation}
\frac{\partial _{g}T_{c}(g)}{T_{c}(g)}{|_{g=0}}=\frac{I_{1}(d,\beta )}{%
k_{B}T_{c}^{0}\lambda _{T_{c}^{0}}^{3}}  \label{deltaTsuT2}
\end{equation}%
where 
\begin{equation}
I_{1}=2\frac{\int d\Sigma \Lambda ^{\prime }\left[ u^{2}+v^{2}(1+dv^{\beta })%
\right] Q\left[ v^{2}(1+dv^{\beta })\right] }{\int d\Sigma \left(
u^{2}+v^{2}(1+dv^{\beta })\right) \Lambda ^{\prime }\left[
u^{2}+v^{2}(1+dv^{\beta })\right] }.  \label{I1}
\end{equation}%
This integral can be evaluated numerically for $b_{1}$ which gives%
\begin{equation}
b_{1}(d,\beta )=2I_{1}(d,\beta ).
\end{equation}%
Therefore one can find a range of values of $d$ and $\beta $ which are in
agreement with the empirical value $b_{1}\simeq -3.5\pm 0.3$ found in Ref.%
\cite{smith}.
\begin{table}[htdp]
\caption{Values of $b_1(\beta,d)$,  $b_2(\beta,d)$ \\
obtained from the parameters $d$ and $\beta$}
\label{tab:1}
\vspace*{.2cm}
\begin{tabular}{llll}
\hline\noalign{\smallskip}
$\beta$& $d$ &$b_1(d,\beta)$&$b_2(d,\beta)$ \\
\noalign{\smallskip}\hline\noalign{\smallskip}
$-1$& $0.01$& -3.41931&18.6006\\
$-1$& $0.1$& -3.36182&17.3356\\
$-1$& $10$& -2.36313&6.64378\\
\noalign{\smallskip}\hline\noalign{\smallskip}
$0$& $0$& -3.42603&18.7765\\
$0$& $0.1$& -3.42603&18.7765\\
$0$ & 1 &-3.42603 &18.7765\\
$0$ & $10$&-3.42603&18.7765 \\
\hline\noalign{\smallskip}
$1$& $0.1$& -3.51504 &20.2565\\
$1$ & 1 & -3.76418 &25.2715\\
$1$ & $10$&-3.97423&31.7773 \\
\hline\noalign{\smallskip}
$2$ & $0.1$ & -3.63134&22.0627 \\
$2$ & $1$ & -3.98266&29.4989\\
$2$ & $10$ &-4.26837&39.7218 \\
\hline\noalign{\smallskip}
\end{tabular}
\vspace*{0cm}  
\end{table}
On the other hand, we may calculate $b_{2}(d,\beta )$ from \label{deltaTsuT2}
for the parameters under consideration from%
\begin{equation}
\begin{array}{lll}
I_{2}(d,\beta )=4\int d\Sigma \left[ \Lambda ^{\prime }\left[
u^{2}+v^{2}(1+dv^{\beta })\right] S\left[ v^{2}(1+dv^{\beta })\right] \right.
&  &  \\ 
+\Lambda ^{\prime \prime }\left[ u^{2}+v^{2}(1+dv^{\beta })\right] \times & 
&  \\ 
\left. \left[ Q\left[ v^{2}(1+dv^{\beta })\right] -{\frac{1}{2}}%
[u^{2}+v^{2}(1+dv^{\beta })]I_{1}(d,\beta )\right] ^{2}\right] / &  &  \\ 
\int d\Sigma \left( u^{2}+v^{2}(1+dv^{\beta })\right) \Lambda ^{\prime } 
\left[ u^{2}+v^{2}(1+dv^{\beta })\right] &  &  \\ 
&  & 
\end{array}
\label{I2b}
\end{equation}%
where 
\begin{equation}
S[\alpha ]\equiv \frac{3}{2}I_{1}(d,\beta )Q[\alpha ]+\left( \alpha
I_{1}(d,\beta )-2Q[\alpha ]\right) g_{1/2}\left[ \exp (-\alpha )\right] .
\label{Sdefinition}
\end{equation}%
From this one obtains%
\begin{equation}
b_{2}(d,\beta )=2I_{2}(d,\beta ).  \label{b2}
\end{equation}%
We remark that the case $\beta =-1$ corresponds to the potential suggested
in Ref.\cite{malik}. Table I shows the results obtained for $b_{1}$ and $%
b_{2}$ from different values of parameters $d$ and $\beta .$ We found that
for $\beta =1$ and $d=1$, $b_{1}\simeq -3.764$ which is in agreement with
the experimental value $b_{1}^{exp}\simeq -3.5\pm 0.3$ obtained in Ref.\cite%
{smith}. We also obtain $b_{2}\simeq 25.27$ which improves upon the result $%
b_{2}\simeq 18.8$ obtained in Ref.\cite{fabio}. However, our estimation for
the parameter $b_{2}$ still remains smaller than the experimental estimation 
$b_{2}^{exp}\simeq 46\pm 5$ reported in Ref. \cite{smith}


\subsection{Temperature-dependent generic potential with free parameter $%
\protect\delta$}

On the other hand, for the potential (\ref{pot2}) Eq. (\ref{Shift}) is only
a function of $\delta $ since 
\begin{equation}
\frac{\partial _{g}T_{c}(g)}{T_{c}(g)}{\Huge |}{_{g=0}}=\frac{I_{1}(\delta )%
}{k_{B}T_{c}^{0}\lambda _{T_{c}^{0}}^{3}}  \label{deltaTsuT2}
\end{equation}%
where now 
\begin{equation}
I_{1}=2\frac{\int d\Sigma \Lambda ^{\prime }\left[ u^{2}+v^{2+\delta }\right]
Q\left[ v^{2+\delta }\right] }{\int d\Sigma \left( u^{2}+v^{2+\delta
}\right) \Lambda ^{\prime }\left[ u^{2}+v^{2+\delta }\right] }.  \label{I1}
\end{equation}%
This integral must also be evaluated numerically in order to obtain the
value of $b_{1}$ 
\begin{equation}
b_{1}(\delta )=2I_{1}(\delta ).
\end{equation}%
Thus, one can find a range of values of $\delta $ which are in agreement
with the empirical value $b_{1}\simeq -3.5\pm 0.3$. Table II shows the
results obtained for $b_{1}(\delta )$ and $b_{2}(\delta )$ from different
values of the parameter $\delta $, we found that, for $\delta =0.5$, $%
b_{1}\simeq -3.7862$ which is in agreement with the experimental value $%
b_{1}^{exp}\simeq -3.5\pm 0.3$ obtained in Ref.\cite{smith}, and
consequently we obtain $b_{2}\simeq 25.986$. %
%
\begin{table}[htdp]
\caption{Values of $b_1(\delta)$ and $b_2(\delta)$ \\
obtained from the parameter $\delta$ }
\label{tab:1}
\vspace*{.2cm}
\begin{tabular}{lll}
\hline\noalign{\smallskip}
$\delta$ &$b_1(\delta)$&$b_2(\delta)$ \\
\noalign{\smallskip}\hline\noalign{\smallskip}
$-0.1$& -3.34203&17.3782\\
 $0$& -3.42603&18.7765\\
 $0.1$& -3.50564&20.1912\\
0.2&-3.58118 &21.621\\
 $0.3$&-3.65295&23.0644 \\
  $0.5$&-3.78626&25.986 \\
   $1$&-4.06981&33.3811 \\
\hline\noalign{\smallskip}
\end{tabular}
\vspace*{0cm}  
\end{table}
\begin{equation}
b_{2}(\delta )=2I_{2}(\delta ).  \label{b2}
\end{equation}%
A similar procedure leads one to%
\begin{equation}
\begin{array}{lll}
I_{2}(\delta )=4\int d\Sigma \left[ \Lambda ^{\prime }\left[
u^{2}+v^{2+\delta }\right] S\left[ v^{2+\delta }\right] +\Lambda ^{\prime
\prime }\left[ u^{2}+v^{2+\delta }\right] \times \right. &  &  \\ 
\left. \left[ Q\left[ v^{2+\delta }\right] -{\frac{1}{2}}(u^{2}+v^{2+\delta
})I_{1}(\delta )\right] ^{2}\right] / &  &  \\ 
\int d\Sigma \left( u^{2}+v^{2+\delta }\right) \Lambda ^{\prime }\left[
u^{2}+v^{2+\delta }\right] &  &  \\ 
&  & 
\end{array}
\label{I2a}
\end{equation}%
\begin{equation}
S[\alpha ]\equiv \frac{3}{2}I_{1}(\delta )Q[\alpha ]+\left( \alpha
I_{1}(\delta )-2Q[\alpha ]\right) g_{1/2}\left[ \exp (-\alpha )\right] .
\label{Sdefinition}
\end{equation}%
from which one obtains $b_{2}(\delta )$ (see Table II).


\section{Conclusions}

We have explored the shift in the condensation temperature up to second
order in the $S$-wave scattering length, for a Bose-Einstein condensate
trapped in a temperature-dependent generic potential, with no further
assumptions that the semiclassical and Hartree-Fock approximations. Using
these facts, we have recovered the usual value for the parameter $b_{1}$,
and consequently, were able to improve the numerical value associated with
the second parameter $b_{2}$ up to $25.271$ for the corresponding potential (%
\ref{pot1}), and $25.986$ for the second potential (\ref{pot2}) compared to
the value obtained in Ref.\cite{fabio} under typical laboratory conditions.
However, the corresponding values for $b_{2}$ obtained here remain smaller
than the experimental value reported in Ref.\cite{smith}. Such disagreement
might be related to effects beyond the HF MF framework or even to
finite-size corrections. Finally, we stress here that the use of
temperature-dependent traps open up a very interesting line of research for
other relevant properties associated with Bose-Einstein condensates.


\begin{acknowledgements}
F.B. is a Marie Curie fellow of the Istituto Nazionale di Alta
Matematica Francesco Severi, MG thanks PAPIIT-IN116911, and MdeLl thanks PAPIIT for grant 100314.
\end{acknowledgements}

\end{document}